\documentclass[%
reprint,
superscriptaddress,
amsmath,
amssymb,
aps,
prl,
floatfix,
]{revtex4-2}

\usepackage{braket}
\usepackage{printlen} % to determine the dimensions of your document
\usepackage{amsthm}
\usepackage{isotope}
\usepackage[
    per-mode=symbol,
    retain-zero-uncertainty=true
    ]{siunitx}
\DeclareSIUnit{\gauss}{G}
\DeclareSIUnit{\phonons}{phonons}
\usepackage{braket}
\usepackage{color}

\usepackage{graphicx}% Include figure files
\graphicspath{{graphics/}}

\usepackage{dcolumn}% Align table columns on decimal point
\usepackage{bm}% bold math

\usepackage{hyperref}
\hypersetup{
pdftitle={Modular variable laser cooling for efficient entropy extraction},
pdfauthor={Segtrap Team},
pdfkeywords = {Ion trapping, Laser cooling},
colorlinks=true, linkcolor=magenta, citecolor=magenta, filecolor=[RGB]{0,0,0}, urlcolor=[RGB]{0,0,0}}

\usepackage{bm}
\usepackage{enumerate}

\usepackage{tikz}
\usepackage{dsfont}

\usetikzlibrary{quantikz}

\usepackage{lipsum} 

\usepackage{subcaption}

\usepackage{ragged2e}
\DeclareCaptionJustification{justified}{\justifying}
\usepackage{printlen}

\usepackage{etoolbox}

\makeatletter
\def\maketitle{
\@author@finish
\title@column\titleblock@produce
\suppressfloats[t]}
\makeatother

\newcounter{PRLsections}

\newcounter{PRLsubsections}[section]

\usepackage{xpatch}
\xpretocmd{\section}{\setcounter{PRLsubsections}{0}}{}{}

\usepackage{wasysym}

\usepackage[innercaption]{sidecap}
\sidecaptionvpos{figure}{c}

\DeclareSymbolFont{yhlargesymbols}{OMX}{yhex}{m}{n} 
\DeclareMathAccent{\widehat}{\mathord}{yhlargesymbols}{"62}

\usepackage{multirow}
\usepackage{array}
\newcolumntype{L}[1]{>{\raggedright\let\newline\\\arraybackslash\hspace{0pt}}m{#1}}
\newcolumntype{C}[1]{>{\centering\let\newline\\\arraybackslash\hspace{0pt}}m{#1}}
\newcolumntype{R}[1]{>{\raggedleft\let\newline\\\arraybackslash\hspace{0pt}}m{#1}}

\usepackage{xspace}

\newtoggle{arXiv}
\toggletrue{arXiv}      % you DO want to see it
\newcommand{\Uni}[2]{\ensuremath{{#1} \mathrm{\:{#2}} }}

\newcommand{\caf}{\ensuremath{^{40}{\textrm{Ca}}^{+}\, }}
\let\oldcite\cite
\renewcommand{\cite}[1]{\mbox{\oldcite{#1}}}

\usepackage{tikz}
\usetikzlibrary{arrows,calc,matrix,fit,positioning}
\usetikzlibrary{decorations.pathreplacing}
\usetikzlibrary{decorations.pathmorphing}
\usetikzlibrary{decorations.markings}
\usetikzlibrary{shapes.geometric}
\usetikzlibrary{decorations.pathreplacing}
\usetikzlibrary{calc}

\newcommand{\expect}[1]{\langle #1 \rangle}

\begin{document}

\title{Modular variable laser cooling for efficient entropy extraction}

\author{B.~de Neeve}
\email[Email: ]{brennan.mn@proton.me}
\affiliation{Institute for Quantum Electronics, ETH Z\"urich, Otto-Stern-Weg 1, 8093 Z\"urich, Switzerland}
\author{T.-L.~Nguyen}
\altaffiliation{Current address: QuantX Labs Pty Ltd, Adelaide, Australia}
\affiliation{Institute for Quantum Electronics, ETH Z\"urich, Otto-Stern-Weg 1, 8093 Z\"urich, Switzerland}
\author{A.~Ferk}
\affiliation{Institute for Quantum Electronics, ETH Z\"urich, Otto-Stern-Weg 1, 8093 Z\"urich, Switzerland}
\author{T.~Behrle}
\affiliation{Institute for Quantum Electronics, ETH Z\"urich, Otto-Stern-Weg 1, 8093 Z\"urich, Switzerland}
\author{F.~Lancellotti}
\affiliation{Institute for Quantum Electronics, ETH Z\"urich, Otto-Stern-Weg 1, 8093 Z\"urich, Switzerland}
\author{M.~Simoni}
\affiliation{Institute for Quantum Electronics, ETH Z\"urich, Otto-Stern-Weg 1, 8093 Z\"urich, Switzerland}
\author{S.~Welte}
\email[Email: ]{stephan.welte@pi5.uni-stuttgart.de}
\altaffiliation{Current address:
5. Physikalisches Institut and Carl-Zeiss-Stiftung Center for Quantum
Photonics Jena - Stuttgart - Ulm, Universität Stuttgart, Pfaffenwaldring
57, 70569 Stuttgart, Germany
}
\affiliation{Institute for Quantum Electronics, ETH Z\"urich, Otto-Stern-Weg 1, 8093 Z\"urich, Switzerland}
\author{J. P. Home}
\email[Corresponding author, Email: ]{jhome@phys.ethz.ch}
\affiliation{Institute for Quantum Electronics, ETH Z\"urich, Otto-Stern-Weg 1, 8093 Z\"urich, Switzerland}
\affiliation{Quantum Center, ETH Z{\"u}rich, 8093 Z{\"u}rich, Switzerland}

\date{\today}

\begin{abstract}
  We propose and experimentally demonstrate a method for laser cooling an oscillator based
  on sequences of spin-state-dependent displacements followed by spin repumping. For a
  thermal state with mean occupation $\bar{n}\gg 1$ the method attains a reduction to
  0.632 of the initial thermal oscillator occupation for two repumps of the two-level spin
  state. This is within a factor of 2.53 of the optimum that might be expected due to the
  reduction of the oscillator entropy by $2 \ln(2)$. We show that the method, which is
  based on encoding the value of the modular-variable of the oscillator into the spin, has
  a simple semi-classical description in terms of a Bayesian update. We demonstrate the
  method experimentally using the internal and motional states of a single trapped ion.
\end{abstract}


\maketitle

\noindent Laser cooling has been transformative in its effect on atomic physics
\cite{Wineland1978, Phillips1998}, and the techniques involved have now spread to
manufactured opto-mechanical systems \cite{Chan2011, 2020delic, 2014aspelmeyer} as well as
to molecules \cite{Shuman2010}. It involves coupling the motional degree of freedom to be
cooled to an ancillary system (often the internal states of an atom) which then releases a
photon into the low occupancy reservoir of the optical vacuum, transferring entropy into
the environment. The purification of a fully mixed two-level spin state results in a
reduction of von Neumann entropy corresponding to the value of $\ln(2)$. However, standard
methods such as Doppler and sideband cooling, devised based on experimental simplicity,
are far from attaining such performance. It is therefore tempting to consider whether
cooling techniques approaching this optimal entropy reduction can exist. Such methods
could have a considerable advantage over current methods for cooling a highly excited
oscillator. For instance, for a thermally occupied oscillator with $\bar{n}\gg 1$ the
entropy scales as $\ln(\bar{n})$ and a reduciton of $\ln(2)$ results in halving the mean
thermal energy.

\begin{figure}[t!]
  \begin{center}
    \includegraphics[width=0.45\textwidth]{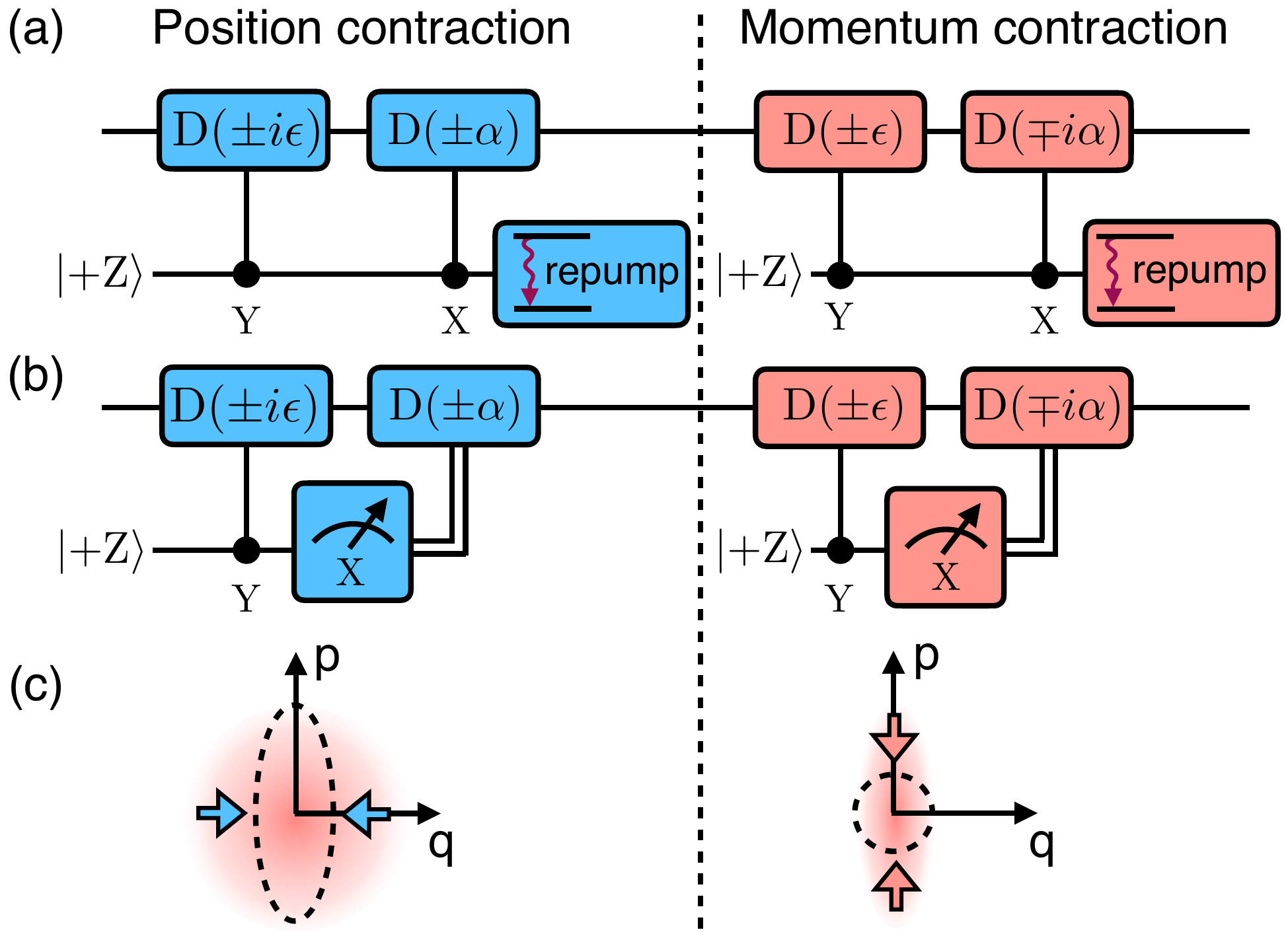}
  \end{center}
  \caption{a) Circuit for a single round of modular-variable cooling of an oscillator.
    Prior to each qubit reset are two state-dependent displacements in orthogonal
    directions and with orthogonal spin control bases. The spin basis used for the
    conditional unitaries is labelled at the control point. This conditions the sign
    (i.e. direction) of the displacement. b) Circuit which is equivalent in its action on
    the oscillator (when we average over measurement outcomes), but which involves
    measurement of the spin and conditional classical feedback. The measurement basis is
    indicated in the measurement symbol. c) Schematic representation of position and
    momentum contraction in phase space acting on an initially thermal state shown by its
    Wigner function.}
  \label{fig:CircuitCool}
\end{figure}

In this Letter, we propose and demonstrate experimentally a method which, unlike standard
cooling methods, reduces entropy in proportion to the fundamental bound given by spin
relaxation. It involves two sequential unitary operations which couple the oscillator to
an ancilliary two-level spin (i.e.\ qubit), which is then reset. The protocol allows the
energy of the thermal oscillator state to be reduced by a factor of 0.632 for each pair of
optical pumping events, for any thermal equilibrium state with $\bar{n} \gg 1$. This is a
constant factor of 2.53 larger than the ultimate limit of $0.25$ energy reduction
attainable by two spin resets (removing a maximum entropy of $2\ln(2)$).

The method, which is inspired by recently demonstrated techniques for stabilizing GKP
states \cite{2001gottesman} in superconducting circuits \cite{CampagneIbarcq2020} and
trapped ions \cite{deNeeve2022}, is based on the circuit shown in
Fig.~\ref{fig:CircuitCool} a), consisting of two spin-controlled oscillator displacements
$U_{\rm CD}(\gamma, \mathcal{P}) \equiv D(\gamma \mathcal{P}) = \exp{(2i \mathcal{P}
  (\mathrm{Im}\{\gamma\}\hat{q} - \mathrm{Re}\{\gamma\}\hat{p}))}$ of amplitude
$\gamma \in \mathbb{C}$, with sign controlled by the internal state of the spin in the
basis of the chosen Pauli operator $\mathcal{P} \in \{X,Y,Z\}$, followed by spin reset by
optical pumping. We have defined the dimensionless phase-space position as
$\hat{q}\equiv \sqrt{m \omega/(2 \hbar)}\hat{q}_r$ and momentum as
$\hat{p} \equiv \sqrt{1/(2 \hbar \omega m)} \hat{p}_r$ and $\hat{q}_r$, $\hat{p}_r$ are
operators for the real space co-ordinates (for this definition
$\left[\hat{q},\hat{p} \right] = i/2$). State-dependent displacements are extensively used
in ion-trap physics \cite{18Fluhmann, Lo2015, Benhelm2008}, while analogous operations
have also recently been realized in superconducting circuits
\cite{CampagneIbarcq2020}. The first of each pair of operations in
Fig.~\ref{fig:CircuitCool} a), which we designate as the \emph{measurement operation},
realizes $U_{\rm CD}(\epsilon \mathrm{e}^{i\theta_m}, Y)$, $\epsilon \in \mathbb{R}$. The
second \emph{correction operation}, realizes
$U_{\rm CD}(\alpha \mathrm{e}^{i(\theta_m-\pi/2)}, X)$, $\alpha \in \mathbb{R}$, and thus
is aligned with the perpendicular quadrature, and acts in the perpendicular spin
basis. For the \emph{first} pair of operations in Fig.~\ref{fig:CircuitCool}~a),
$\theta_m = \pi/2$ and the measurement displacement is aligned with the imaginary
phase-space axis, with the corresponding correction a displacement along the real
phase-space axis. Following the unitary action,
$U_q = U_{\rm CD}(\alpha, X) U_{\rm CD}(i \epsilon, Y)$, of these two
controlled-displacements, the qubit state is reset by optical pumping. We refer to this as
a position contraction. The same sequence is then performed, but with $\theta_m = 0$,
realizing a unitary $U_p$ which after reset results in a momentum contraction.  Assuming
for the moment no effect on the oscillator during qubit reset, and denoting the $\pm 1$
eigenstates of Pauli operators by $\ket{\pm \mathcal{P}}$ the output density matrix for an
initial oscillator state $\rho_j$ can be obtained using the Kraus representation as
\begin{equation}
	\label{eq:rho}
	\rho_{j + 1} = \sum_{\nu, \mu = \pm} K_{p, \nu} K_{q, \mu} \rho_j K_{q, \mu}^{\dag} K_{p, \nu}^{\dag} \, ,
\end{equation}
with the corresponding Kraus operators $K_{q, \pm} \equiv \bra{\pm X} U_q \ket{+Z}$
\begin{align}
  K_{q, \pm} = \mathrm{e}^{\mp 2i \alpha \hat{p}} \cos\left( 2\epsilon \hat{q} \pm \pi/4 \right) \, .
  \label{eq:Krausx}
\end{align}
and a similar form obtained from the momentum contraction as
$K_{p, \pm} \equiv \bra{\pm X} U_p \ket{+Z}$.

To understand the action of this operation on the oscillator state, it is useful to
consider the modified sequence of Fig.~\ref{fig:CircuitCool}b), which produces the same
Kraus map, but involves a modular value measurement followed by conditional feedback
\cite{CampagneIbarcq2020, 18Fluhmann}. The measurement probabilities are given by
$P({\pm X} | \rho) = \frac{1}{2}(1 \mp \mathrm{Tr}(\hat{O}\rho))$ with
${\hat{O} = \sin\left(4 \epsilon \hat{q} \right)}$ and $\rho$ the initial motional density
matrix. For an initial position eigenstate satisfying $\hat{q}\ket{q} = q \ket{q}$, we
then obtain a conditional probability
\begin{equation}
  P(\pm X | q) = \frac{1}{2} \left(1 \mp  \sin(4 \epsilon q) \right) \, .
  \label{eq:condprob}
\end{equation}
In the Bloch-sphere picture, we can view the spin state prior to the measurement as having
been rotated about the $Y$ axis by an angle $-4 \epsilon q$. For more general states the
spin state will be a superposition corresponding to the distribution of the oscillator in
$q$. The projective spin state measurement result is used to condition the direction of
the feedback displacement which should be applied to the oscillator. For cooling this
should be chosen such as to move the state towards the phase-space origin on average, as
will be explained below.
% The result of this protocol when we average over measurement outcomes is identical to
% that in the initial circuit, except that in the latter the feedback is performed
% coherently before the measurement, and the measurement result is discarded.

\begin{figure}[tb]
\begin{center}
\includegraphics[width=0.43\textwidth]{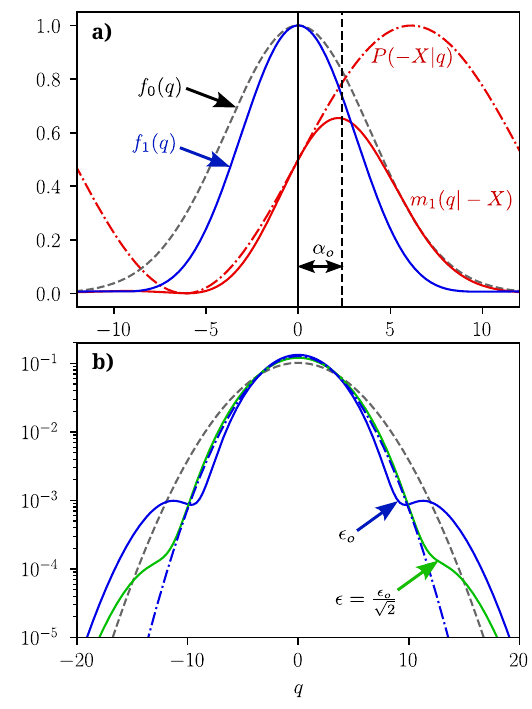}
\end{center}
\caption{a) Position contraction in the semi-classical Bayes description. The initial
  Gaussian probability density (grey dashed curve) is normalized to a peak height of
  unity. The modular measurement produces a conditional probability $P(-X|q)$ given here
  for the optimal value $\epsilon_o$ (given by the dot-dashed red curve), producing the
  modified conditional probability $m_1(q|{-X})$ (solid red curve) which is close to a
  Gaussian displaced from the origin by $\alpha_o$. The narrowed post-correction
  probability density is $f_1(q)$ (solid blue curve), again normalized to a peak height of
  1. b) Probability densities for optimal post-contraction settings (blue solid) and with
  $\epsilon = \epsilon_o/\sqrt{2}$ (green solid) on a logarithmic scale, emphasizing the
  deviations from a Gaussian distribution. The latter produces a smaller reduction in the
  mean energy but reduces the ``tails'' of the distribution. Additional curves show
  Gaussian distributions for both the initial temperature (grey dashed), and for the
  optimal temperature after one round of cooling (blue dot-dashed).}
\label{fig:SemiClassical}
\end{figure}
This interpretation of the quantum mechanical operation in terms of a measurement plus
conditional feedback motivates an understanding through the following semi-classical
treatment, which assumes the oscillator to be a classical variable and can be understood
in terms of a Bayesian update of a probability density. We assume that the initial state
is thermal, with a Gaussian joint probability density $w_0(q, p) = f_0(q) f_0(p)$ with
$f_0(q) = 1/(s \sqrt{2\pi})\exp(- q^2/(2 s^2))$. For the co-ordinate definitions above,
the relation to the thermal occupancy is $s = \sqrt{\left(\bar{n} + 1/2\right)/2}$. Let us
first consider the reduction in the position quadrature. The modular variable measurement
produces a conditional probability Eq.~\ref{eq:condprob} which allows us to update the
probability density of the oscillator according to Bayes rule as
$m_1(q|{\pm X}) \propto P({\pm X} | q) f_0(q)$. Fig.~\ref{fig:SemiClassical} a) shows
$f_0(q)$, $P({-X} | q)$ and $m_1(q | {-X})$. For the parameter choice made in this
example, $m_1(q | {-X})$ approximates a Gaussian distribution which is narrower than
$f_0(q)$ with its center displaced from the origin. To recover an approximate thermal
distribution, a displacement of the probability density back to the origin is required,
which is accomplished by the correction displacement operation
${\hat{T}_{\mathrm{class}}(q, -\alpha): f(q) \mapsto f(q + \alpha)}$. Similar logic holds
conditioned on the measurement result $+X$, with the opposite sign on the correction
displacement. Summing over these two possibilities, the probability density in position
becomes $f_1(q) = P(+X) m_1({q-\alpha}|{+X}) + P(-X) m_1({q + \alpha}|{-X})$. Since the
position and momentum are classically independent variables, the same procedure can be
repeated using appropriate settings for the momentum quadrature, leading to a final state
described classically by ${w_1(q,p) = f_1(q) f_1(p)}$. The expectation value of the energy
for $w_1(q,p)$ is
\begin{equation}
  \label{eq:meaneclass}
  \langle E \rangle_{\mathrm{classical}} =  2 \hbar \omega
  \left(\alpha ^2 + s^2 \left(1 - 8 \alpha \epsilon  \mathrm{e}^{-8 \epsilon ^2 s^2}
    \right)\right) \, .
\end{equation}
This expression can be minimized with respect to $\alpha, \epsilon$ to find the most
effective settings for cooling in the limit where the classical approximation is valid,
resulting in the optimal parameter values $\alpha_o = s/\sqrt{\mathrm{e}}$,
$\epsilon_o = 1/(4s)$ and a reduction in the energy expectation value by a factor
$(\mathrm{e}-1)/\mathrm{e} \simeq 0.632$. In Fig.~\ref{fig:SemiClassical} b) we show the
phase-space distribution in one quadrature using these values on a log scale, which
emphasizes that $w_1(q,p)$ is not an exact Gaussian but is a close approximation. The
issue of the residual tails in the distribution will be discussed below.

% The corresponding mean energy can be obtained for a thermal initial state by first
% calculating the characteristic function
% $\chi(\beta) = \mathrm{Tr}(D(\beta) \rho_{j + 1})$ and then performing the appropriate
% derivatives \cite{2002barnett}.
A more exact result can be obtained from the density matrix $\rho_1$ obtained from
Eq.~\ref{eq:rho} with $\rho_0$ a thermal state, giving
\begin{align}
  \label{eq:quant-energy}
  &\expect{E}_{\mathrm{quantum}} =
    \expect{E}_{\mathrm{classical}} + \\
  & 2\hbar\omega \bigg( \epsilon^2 - \alpha\epsilon\mathrm{e}^{-8\epsilon^2 s^2}
    \Big( 4s^2(\cos(4\epsilon^2)-1) + \sin(4\epsilon^2) \Big) \bigg) \, .
    \nonumber
\end{align}
where $\expect{E}_{\mathrm{classical}}$ is obtained from Eq.~\ref{eq:meaneclass}. Close to
the optimal values $\alpha_o, \epsilon_o$ the difference between the classical and quantum
result is of order $1/s^2$, and is thus a minor correction for $\bar{n} \gg 1$. As the
system approaches the ground state, the correction to the classical result becomes more
significant, and the optimal parameters deviate from $\alpha_{o},
\epsilon_{o}$. Nevertheless, we observe that it is possible to find values of $\epsilon$
and $\alpha$ which produce cooling well below $\bar{n} = 1$. For a given value of
$\epsilon$ the energy $\expect{E}_{\mathrm{quantum}}$ is minimized by choosing
\begin{equation}
  \label{eq:opt-alpha}
  \alpha = \frac{1}{2}\epsilon \mathrm{e}^{-8\epsilon^2 s^2}
  \left( 4s^2\left( 1 + \cos\left(4\epsilon^2\right) \right)
    + \sin\left(4\epsilon^2\right) \right) \, .
\end{equation}
We can find the optimal value of $\epsilon$ by setting $\alpha$ as in
Eq.~\ref{eq:opt-alpha} and minimizing Eq.~\ref{eq:quant-energy} numerically.

We demonstrate the method using the axial motion of a single trapped \caf ion with a
frequency of around $\omega_m \approx 2 \pi \times \Uni{1.7}{MHz}$. The motional mode is
controlled and read out via the internal electronic levels
$\ket{+Z}_s \equiv \ket{^{2}{S}_{1/2}, m_j=1/2}$ and
$\ket{-Z}_s \equiv \ket{^{2}{D}_{5/2}, m_j=3/2}$, separated by a transition at 729 nm,
forming the internal-state qubit (where subscript $s$ refers to \emph{spin}). The
electronic state is initially prepared in $\ket{+Z}_s$. Then the two state-dependent
displacements are implemented by application of a bi-chromatic laser pulse simultaneously
driving the red and blue motional sidebands of the qubit transition, realizing the
Hamiltonian
$ \hat{H} = \hbar \eta \Omega \sigma_{\phi_s} \left(\sin(\phi_m)\hat{q} -
  \cos(\phi_m)\hat{p}\right) $ with
$\sigma_{\phi_s} \equiv \left(\cos(\phi_s) X + \sin(\phi_s) Y \right)$ and
$\eta \simeq 0.05$ the Lamb-Dicke parameter~\cite{98Wineland2}.
% When this is applied for a finite time $t$, it results in an operation
% $\exp( \sigma_\phi (\chi\create - \chi^* \destroy) )$ with
% $\chi = i\Omega t e^{i\phi_m}/2$.
The spin phase $\phi_s$ can be chosen by the mean phase of the two laser drives, and the
motional phase $\phi_m$ by the difference of their phases. Applied for a time period $t$
this realizes $U_{\mathrm{CD}}(\gamma, \mathcal{P})$ with
$\gamma(t) = \mp \eta\Omega t e^{\phi_m}/2$, allowing $\Omega t$ and $\phi_m$ to be used
to control $\epsilon$, $\alpha$, $\theta_m$, and $\theta_c$. Reset of the qubit is
performed by optical pumping, which we estimate requires scattering an average close to 2
photons, one at 854 nm and one at 397 nm \cite{deNeeve2022}.

We prepare initial thermal states of the oscillator with $\bar{n}$ of $14.7 \pm 0.8$,
$34.0 \pm 1.8$, and $51.1 \pm 4.4$ by performing Doppler cooling with selected laser
powers and detunings. We characterize the temperature of these initial states by measuring
the characteristic function using state-dependent displacements \cite{2020fluhmann}, and
fitting the obtained data with the Gaussian form expected for a thermally excited
oscillator. For each initial state we extract the value of $\bar{n}$ from the width of the
Gaussian, which is narrower for larger $\bar{n}$. At higher occupations, we find that to
obtain accurate estimates of the values of $\bar{n}$, we must include higher order
interactions in the Hamiltonian that arise outside the Lamb-Dicke regime
\cite{98Wineland2}, including the effects of thermally excited radial modes with
approximate frequencies of $2 \pi \times \Uni{2.4}{MHz}$ and
$2 \pi \times \Uni{3.2}{MHz}$. The latter modify the coupling rate $\Omega$. Further
details on the estimation of $\bar{n}$ from characteristic function readout are given in
the Supplemental Material.

We then apply the modular variable cooling for several rounds and subsequently measure the
characteristic function of the oscillator, from which we extract $\bar{n}$ as described
above for thermal states. The values of $\alpha$ and $\epsilon$ at each round were chosen
assuming that $\bar{n}$ is reduced by the optimal amount. Unlike the initial states, the
cooled states are only approximately thermal and contain high energy ``tails'', as noted
above, but the Gaussian fit allows us to obtain the mean occupation of the ``thermal
fraction''.
\begin{figure}[tb]
  \begin{center}
    \includegraphics[width=0.5\textwidth]{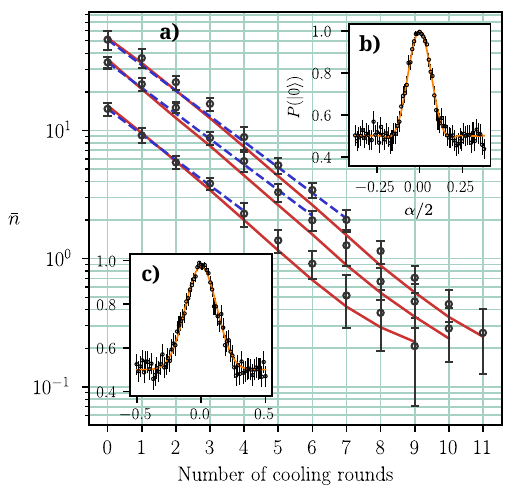}
  \end{center}
  \caption{
    % Experimental results working with a trapped-ion oscillator. Measurements are
    % from 31.8.2022. In these measurements, we chose an initial temperature of
    % $\overline{n}{=}29.4$.
    Estimated values of $\bar{n}$ after each cooling round, starting from initial thermal
    states prepared with ${\bar{n} \approx 15}$, 34, and 51. a) Error bars are $95 \,\%$
    confidence intervals obtained from Gaussian fits of experimentally measured
    characteristic functions. The blue dashed lines give
    $\bar{n}_0 \times 0.632^{\mathrm{rounds}}$ predicted by the semi-classical theory,
    which due to quantum effects is not expected to be valid for ${\bar{n} < 2}$. Red
    lines give $\bar{n}$ predicted by numerical simulations of the experiment including
    several known imperfections. The insets show fits to characteristic functions readout
    data for b) an initial state with $\bar{n} \approx 34$ and c) a state after three
    cooling rounds starting from initial state with $\bar{n} \approx 34$, for which the
    obtained $\bar{n} \approx 8.5$.}
  \label{fig:ExpResults}
\end{figure}
%% For values of $\bar{n} > 2$ we additionally plot the $\bar{n}$ reduction of 0.632
%% expected from the semi-classical analysis above. For lower $\bar{n}$ we expect that the
%% deviation
% from the semi-classical treatment can no longer be neglected. We also plot the results
% of simulations where we account for many imperfections that we have in the experiment
% (see Supplemental Material). We extract the plotted $\bar{n}$ values from the density
% matrix in the simulation by also simulating the characteristic function readout and
% fitting to a thermal distribution as done for the experiment.
The values thus obtained are plotted in Fig.~\ref{fig:ExpResults}, along with a comparison
to the scaling $0.632^{\mathrm{rounds}}$ as well as numerical simulations which take
account of known experimental imperfections. For the first few rounds of cooling we
observe good agreement between the simulation and semi-classical theory. After more rounds
the simulation deviates from a thermal state which may explain the colder states predicted
compared to the semi-classical theory; although the reduction of $\bar{n}$ of 0.632
predicted by the semi-classical theory makes no thermal assumption on the \emph{final}
state after a \emph{single} round, by assuming a 0.632 reduction over many rounds we are
implicitly assuming the state is thermal before \emph{each} round of cooling is
applied. Thus the values of $\bar{n}$ extracted from the simulation gradually deviate
below the semi-classical theory as the high-energy components are increasingly neglected
by the semi-classical theory. The results from the experiment mostly agree quite well with
the semi-classical theory, but tend to slightly higher values of $\bar{n}$ than the
simulation in the early rounds.

The deviation from a thermal state increases with the temperature of the initial thermal
state and the number of cooling rounds. After performing many rounds of cooling, a large
fraction of the population reaches the ground state, while some remains in the tails of
the distribution at high energy. We quantify the amount of population close to the ground
state by performing a blue-sideband pulse after the cooling and looking at the contrast of
the resulting Rabi oscillations. We have thus found populations of $\sim 2.6 \%$, $7 \%$,
and $13 \%$ remaining at high energy in the final states after cooling from initial
$\bar{n} \approx$ 15, 34, and 51 respectively. Further details are given in the
Supplemental Material. Theoretically we see (Fig.~\ref{fig:SemiClassical} b)) that using a
less optimal value of $\epsilon = \epsilon_o/\sqrt{2}$ for the cooling leads to high
energy tails in the resulting final state that are roughly an order of magnitude lower. We
have tried this experimentally, but have not observed a notable change in the high energy
tails after many rounds of cooling, although the numerical simulations do show a
reduction. The reasons for this are unclear -- a more detailed discussion is given in the
Supplemental Material.

We conclude that the method is particularly effective at cooling high temperature states,
but also limited by the occurrence of high energy tails. Our results show that the highest
temperature states that can be effectively cooled with an acceptably low population
remaining in high energy tails is limited by the noise model of a particular
experiment. We expect that for systems with lower noise and the capability to perform
higher fidelity coherent operations the method will become particularly attractive,
offering the means to remove a large fraction of the oscillator energy with two repumps
while keeping the high energy tails in the final state negligible.

It is instructive to compare the modular variable cooling to Doppler cooling, for an
analogy can be drawn based on the classical treatment by considering that the conditional
excitation probability for the momentum quadrature, $P(\pm Z|p)$ is proportional to the
Lorentzian of the resonance of the transition, while for the position quadrature, the
conditional probability is flat.
% Close to the Doppler limit, the range of velocities of the initial state is much smaller
% than the Lorentzian width, and thus the initial Gaussian distribution sits within a
% region of the Lorentzian for which the response varies approximately linearly as a
% function of momentum \cite{2003leibfried}.
Making the assumption that the range of velocities of the initial state is much smaller
than the width of the Lorentzian, a slightly naive approach is then to match the linear
approximation of the Lorentzian to a linear approximation
$ P(\pm X|p) = \frac{1}{2}\left( 1 \pm \sin\left( 4\epsilon p \right) \right) \approx
\frac{1}{2} \pm 2\epsilon p $, which results in ${\epsilon = \eta \omega/\Gamma}$ with
$\eta$ the Lamb-Dicke parameter ${\eta = \sqrt{E_{\rm recoil}/{\hbar \omega}}}$ and
$\Gamma$ the decay rate of the transition. The correction displacement is given by the
recoil of a single photon, thus is also of size $\eta$. Using Eq.~\ref{eq:meaneclass}, the
average change in energy is then
$2\hbar\omega\eta^2 \left( 1 - 2(\nu/\Gamma)(2\bar{n} + 1)\mathrm{e}^{-2\eta^2 \nu^2
    (2\bar{n} + 1)/\Gamma^2} \right)$. For values such that the exponential factor can be
approximated as 1 (Lamb-Dicke and weak-binding regimes \cite{Wineland79}) this change is
zero for $\bar{n} + 1/2 = \Gamma/(4 \omega)$ which is what would be expected as the
Doppler limit of the standard theory neglecting the recoil in spontaneous emission
\cite{2003leibfried}. Further details are given in the Supplemental Material. For high
$\bar{n}$, Doppler cooling produces a fractional reduction in energy which is far from
optimal, but it is robust and fast because both the gradient and the measurement
correction are far lower than the optimal values found above and the rate of scattering is
very high.

% A second common cooling method is sideband cooling, which is commonly performed using a
% Jaynes-Cummings Hamiltonian. This can be written in the form
% $H_{\rm JC} = \Omega \left(X \hat{q} +Y \hat{p}\right)$. Thus the combination of
% state-dependent displacements used in the protocol of the current paper appear in
% sideband cooling, although they are applied simultaneously and with the same
% magnitude. This constraint is relaxed in the cooling protocol demonstrated in this
% Letter. We note that in trapped-ion experiments the spin repump operation scatters
% photons which induce recoil forces on the ion thereby heating the motional degree of
% freedom. At high $\bar{n}$ the heating due to recoil is negligible, but as we approach
% the ground state the recoil will eventually limit further cooling. In that limit, the
% method is no more efficient than sideband cooling even if we neglect recoil, and has the
% disadvantage that it does not reach a dark steady-state of the unitary interaction, for
% which the spin excitation is suppressed.

The method demonstrated here achieves a reduction in the mean vibrational quantum number
by a factor of $0.632$ for a single round, which requires two spin resets. Fundamental
considerations would imply that the possible reduction in entropy from two spin resets is
$2\ln(2)$. An oscillator thermal state has an entropy of
$- \bar{n} \ln\left(\bar{n}/(\bar{n}+1)\right) + \ln(\bar{n}+1)$, which for $\bar{n}\gg 1$
simplifies to $S_{\mathrm{th}}\simeq \ln(\bar{n})$. Subtracting the entropy from two spin
resets then leaves a result $S_{\mathrm{th}}' \simeq \ln(\bar{n}/4)$, which is consistent
with reduction of $\bar{n}$ by a factor of 4. This implies that in the limit
$\bar{n} \gg 1$ an improvement of up to ${4\times 0.632 = 2.53}$ might be achievable by
further optimization of the pulse sequence used to implement the cooling, which should
form the focus of future investigation. The methods here could be generalized to other
states and platforms. The similarity to the earlier work on the stabilization of grid
states indicates sequences similar to those given above could be used to cool into highly
non-classical states \cite{CampagneIbarcq2020, deNeeve2022}. These first examples point to
new possibilities for quantum state generation and control. Due to the appearance of
similar Hamiltonians, we expect that these methods will find application in systems beyond
trapped ions and atoms, including superconducting circuits and nano-mechanical
oscillators.

JH devised the scheme and JH and BN performed the theoretical study. BN, TL, SW, and TB
carried out the experiments, and AF developed some required control system features and
helped BN to implement the cooling sequence. BN and TL performed numerical
simulations. Experimental assistance was provided by FL and MS. The paper was written by
JH, BN and SW with input from all authors.

We thank P. Campagne-Ibarcq for stimulating discussions. We acknowledge support from the
Swiss National Science Foundation through the National Centre of Competence in Research
for Quantum Science and Technology (QSIT) grant 51NF40–160591, and from the Swiss National
Science Foundation under grant number 200020 165555/1. S.W. acknowledges financial support
via the SNSF Swiss Postdoctoral Fellowship (Project no. TMPFP2$\_$210584).

\bibliography{../myrefsStateTomo}

% \clearpage

\iftoggle{arXiv}{
	\clearpage 
	\title{Supplemental Material: Modular variable laser cooling for efficient entropy extraction}
    \maketitle
}{}

\label{sec:supplement}

% Figure counter and label for the Supplemental Material.
\setcounter{figure}{0}
\renewcommand{\thefigure}{S.\arabic{figure}}

% Equation counter and label for the Supplemental Material.
\setcounter{equation}{0}
\renewcommand{\theequation}{S.\arabic{equation}}

\section{Quantum treatment of cooling process}

Calculations of the update to the energy following a single round of cooling were
performed using the normal ordered characteristic function, which is defined for the
density matrix $\rho$ as
\begin{equation}
  \chi(\beta) = \mathrm{Tr}\left(D(\beta) \rho\right)\mathrm{e}^{|\beta|^2/2} \ .
\end{equation}
The relevant density matrix is given by $\rho_1$ in Eq.~\ref{eq:rho} in the main
manuscript, which consists of a sum of 16 terms weighted by $1/16$ with a complex
amplitude $c_{\gamma, \chi}$, each of which constitutes a displacement on the left (right)
by a particular value $\gamma$ ($\chi$) of the original thermal density matrix
$\rho_0 = \sum_n p_n \ket{n}\bra{n}$, with $\ket{n}$ the energy eigenstates, and
$p_n = \bar{n}^n/(\bar{n}+1)^{n + 1}$ the relevant probabilities for a thermal
distribution. The result for the characteristic function is a sum of terms
\begin{align}
  &\chi_{\rho_1}(\beta) =\\
  &\frac{e^{|\beta|^2/2}}{16}\sum_{\{\gamma, \chi\}}c_{\gamma, \chi}
    \sum_{ n}  p_n \mathrm{Tr}\Big(D(\beta) D(\chi) \ket{n}\bra{n} D(-\gamma)\Big) \nonumber
\end{align}
which using the cyclic property of the trace gives
\begin{equation}
  \chi_{\rho_1}(\beta) = \frac{e^{|\beta|^2/2}}{16}\sum_{\{\Upsilon\}} c_{\gamma, \chi}
  \sum_{n} p_n \bra{n} D(\Upsilon) \ket{n} \ ,
\end{equation}
where $D(\Upsilon){=}D(-\gamma) D(\beta) D(\chi)$. The sum over a thermal distribution of
the diagonal elements of a displacement operator in the energy eigenbasis
\cite{1969cahill} then gives
\begin{equation}
  \chi_{\rho_1}(\beta) = \sum_{\Upsilon} e^{-|\Upsilon|^2 \bar{n}} \ .
\end{equation}
The mean occupation can be found from the characteristic function using
\begin{equation}
  \langle n \rangle =
  -\frac{\partial^2}{\partial \beta \partial \beta^*}\chi_{\rho_1}(\beta) \bigg|_{\beta = 0} \ .
\end{equation}

\section{Characteristic function readout}\label{sec:char-func-readout}

We estimate the values of $\bar{n}$ of initial thermal states in the experiment as well as
approximate thermal states after cooling by applying a state-dependent displacement
$D(\alpha X/2)$. Given an oscillator state density matrix $\rho$ the measurement
probability of the spin is then
\begin{equation}
  P(\ket{+Z}) = \frac{1}{2}\left(1 + \mathrm{Re}\right\{
  \left\langle D(\alpha) \right\rangle \left\} \right) \ ,
  \label{eq:char-func-readout-prob}
\end{equation}
allowing to determine the real part of $\langle D(\alpha) \rangle$, which is the symmetric
ordered characteristic function \cite{2002barnett}. One may also measure the imaginary
part of the characteristic function by performing an appropriate single-qubit rotation
before the measurement \cite{2020fluhmann}. We do not do this because the characteristic
function for thermal states is real-valued. The Hamiltonian
\begin{equation}
  \label{eq:sdfHam}
  \hat{H} = \hbar \eta \Omega \sigma_{\phi_s} \left(\sin(\phi_m)\hat{q} -
    \cos(\phi_m)\hat{p}\right)
\end{equation}
(where $\sigma_{\phi_s} \equiv \left(\cos(\phi_s) X + \sin(\phi_s) Y \right)$ and
$\eta \simeq 0.05$ is the Lamb-Dicke parameter) that is used to realise the
state-dependent displacement $D(\alpha X/2)$ is only the lowest order term of the
Lamb-Dicke expansion \cite{98Wineland2}, and for highly excited thermal states we fit the
readout data using a simulation that models the full expansion. In addition, since the
radial modes are expected to also be in thermal states we must account for their effect,
which is to modify the effective Rabi rate $\Omega$ in Eq.~\ref{eq:sdfHam} dependent on
the occupancy of the radial modes. In particular, if each of the two radial modes are in
energy eigenstates $\ket{n_1}$, $\ket{n_2}$, then the effective Rabi frequency of the
drive is
\begin{equation}
  \Omega\prod_{j=1,2}\mathrm{e}^{-|\eta_j|^2/2} L_{n_j}(|\eta_j|^2) \ ,
  \label{eq:radial-Rabi-Fock}
\end{equation}
where $L_n(x)$ are Laguerre polynomials. For the experiments we performed the radial mode
frequencies were $\omega_1 \simeq 2 \pi \times \Uni{2.4}{MHz}$,
$\omega_2 \simeq 2 \pi \times \Uni{3.2}{MHz}$, and the Lamb-Dicke parameters
$\eta_1 \simeq 0.02$, $\eta_2 \simeq 0.03$. For a thermal state the probability of radial
mode occupancy $n_j$ is $P_{n_j} = \bar{n}_j^{n_j}/(\bar{n}_j + 1)^{n_j+1}$, leading to a
\emph{distribution} of Rabi frequencies given by Eq.~\ref{eq:radial-Rabi-Fock}. We model
the effect by a simulation that samples the Rabi frequency from the thermal distribution
for each radial mode. We then repeat and average the result of the simulation over many
samples. The values of $\bar{n}_j$ are estimated assuming each mode to reach the same
temperature in the steady-state of Doppler cooling.
\begin{figure}[h]
  \centering
  \includegraphics[width=0.475\textwidth]{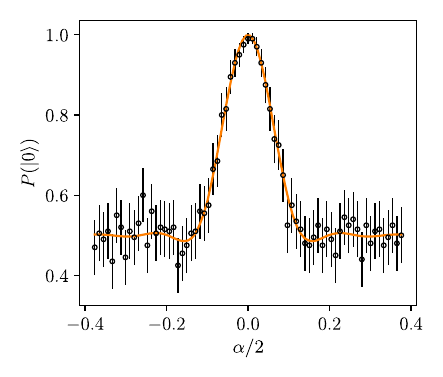}
  \caption{Characteristic function readout of a thermal state with $\bar{n} \approx 51$
    (highest $\bar{n}$ initial state in Fig.~\ref{fig:ExpResults} of the main text). The
    lowest order term in the Lamb-Dicke expansion would lead to readout probabilities
    given by Eq.~\ref{eq:char-func-readout-prob} which is a Gaussian function for thermal
    states. The non-Gaussian features are due to higher order terms in the Lamb-Dicke
    expansion. Each data point is the average over 200 experimental shots. Error bars are
    $95\; \%$ confidence intervals of binomial errors.}
  \label{fig:sdf-readout-fit}
\end{figure}
\begin{figure}[h]
  \centering
  \includegraphics[width=0.5\textwidth]{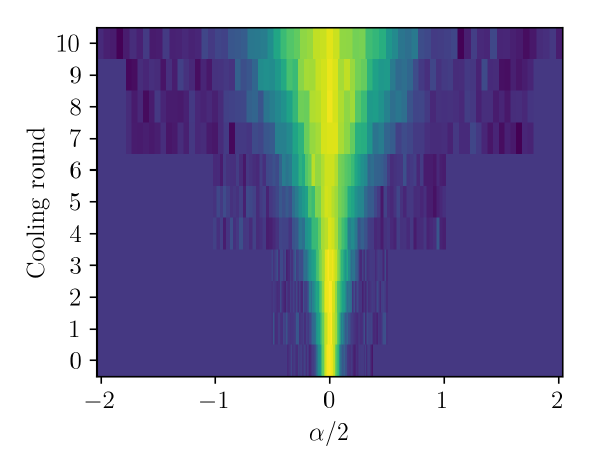}
  \caption{Characteristic function readouts for an initial thermal state with
    ${\bar{n} \approx 34}$ and after each of 10 cooling rounds, corresponding to the
    intermediate temperature data plotted in Fig.~\ref{fig:ExpResults} of the main
    text. The central peak reaches $P(\ket{+Z}) = 1$ and the probability $P(\ket{+Z})$
    drops to $\sim 1/2$ for large values of $\alpha$. Each point is the average of 400
    experimental shots.}
  \label{fig:sdf-readout-rounds}
\end{figure}
In Fig.~\ref{fig:sdf-readout-fit} we show an example plot of characteristic function
readout data for an initially prepared thermal state with $\bar{n} \approx 51$ in the
axial mode with a fit that uses the described model. If the readout probabilities were
given simply by Eq.~\ref{eq:char-func-readout-prob}, the functional form would be
Gaussian, however the higher order terms in the Lamb-Dicke expansion lead to non-Gaussian
features seen in the figure as small oscillations in addition to the overall Gaussian form
expected from a thermal characteristic function. In Fig.~\ref{fig:sdf-readout-rounds} we
plot experimentally measured characteristic function readout data corresponding to the
curve in centre of Fig.~\ref{fig:ExpResults} a), starting from an initial thermal state
with $\bar{n} \approx 34$. The symmetric ordered characteristic function of the initial
thermal state is a narrow Gaussian, and as the state of the oscillator is cooled the
resulting approximate thermal state has an approximately Gaussian characteristic function
which is broader than the initial state. After 10 rounds of cooling most of the population
reaches the ground state having a characteristic function with a large variance (the
symmetric ordered characteristic function is the two-dimensional Fourier transform of the
Wigner function \cite{2002barnett}).

\section{Numerical simulation}

In Fig.~\ref{fig:ExpResults} of the main text we have plotted $\bar{n}$ estimates obtained
from numerical simulations of the cooling process. For these simulations the state of the
spin and oscillator systems is represented by a density matrix, where the oscillator
Hilbert space is truncated in the energy eigenbasis. Both effects mentioned above to
estimate $\bar{n}$ from characteristic function readouts, namely higher order terms in the
Lamb-Dicke expansion and the effects of thermally excited radial modes have been included
in these simulations as well. In addition we have modelled Markovian noise including
\begin{itemize}
\item Spin dephasing with collapse operator $Z$.
\item Oscillator dephasing with collapse operator ${a^\dag a}$.
\item Oscillator heating with collapse operator $a^\dag$.
\end{itemize}
For the spin dephasing we have modelled some non-Markovian noise as well using a term in
the Master equation of the form
\begin{equation}
  \frac{d \rho}{d t} = \frac{g^2}{K} \left( \mathrm{e}^{-K t} -1 \right) [Z,[Z,\rho]] \ ,
\end{equation}
which models both Gaussian (non-Markovian) and exponential (Markovian) decay to better
approximate what we have measured in the experiment. For the spin dephasing we have used
$g = 1/1.6\; \mathrm{kHz}$, and $K = 1/5\; \mathrm{kHz}$ to model a coherence time of
${\sim 1.6\; \mathrm{ms}}$ measured experimentally with a similar component of Gaussian
decay. For the oscillator dephasing we have modelled mainly Markovian dephasing with an
overall coherence time of $\sim 15\; \mathrm{ms}$ as measured in the experiment, and we
model purely Markovian oscillator heating at a rate of 10 quanta per second.

We model photon recoil due to the repump operations as random displacements sampled from a
distribution taking into account the beam geometry and dipole emission pattern. The random
displacements are applied to the component of the spin-oscillator state that is in the
excited state $\ket{-Z}_s$ while the component in $\ket{+Z}_s$ is left unchanged.

We also model 50 Hz (and higher harmonics) oscillations of the frequency of the axial mode
of ion motion that were measured experimentally. Further details can be found in the
Methods of ref. \cite{deNeeve2022}. We also simulate a small amount of Gaussian random
fluctuations on the axial mode frequency with a standard deviation of 20 Hz.

We observe fluctuations in the temperature of the initial state over the timescales of our
measurements. Since we run many trajectories of the Master equation evolution in our
simulations in order to sample radial mode occupancies, we can sample other parameters
with no additional computational cost. We thus model fluctuations of the temperature of
the initial state by sampling a Gaussian distribution of initial $\bar{n}$ values for the
initial thermal state. We have set the standard deviation in $\bar{n}$ of the distribution
from 2 for lower temperature initial state ($\bar{n} \approx 15$) up to 3 for higher
temperature states ($\bar{n} \approx 51$). Similarly, we model fluctuations in calibrated
laser pulse parameters: Rabi frequency, and laser detuning from the internal ion qubit
transition by sampling Gaussian distributions for these parameters for each trajectory.

\section{High energy tails}

As shown graphically in Fig.~\ref{fig:SemiClassical} b) of the main text we expect that
after performing a round of cooling on an initially thermal oscillator state, the final
state is only approximately thermal and contains tails in the position and momentum
distribution of the state. When performing further rounds of cooling we expect that the
tails remaining from each round accumulate. Thus, after many rounds of cooling a large
fraction of the population reaches the ground state while a small fraction remains at
higher energy. We expect the fractional amount of population at higher energy in the final
state as well as the energy of this population to increase with the temperature of the
initial thermal state, before cooling is applied.

\begin{figure}[p]
  \centering
  \includegraphics[width=0.5\textwidth]{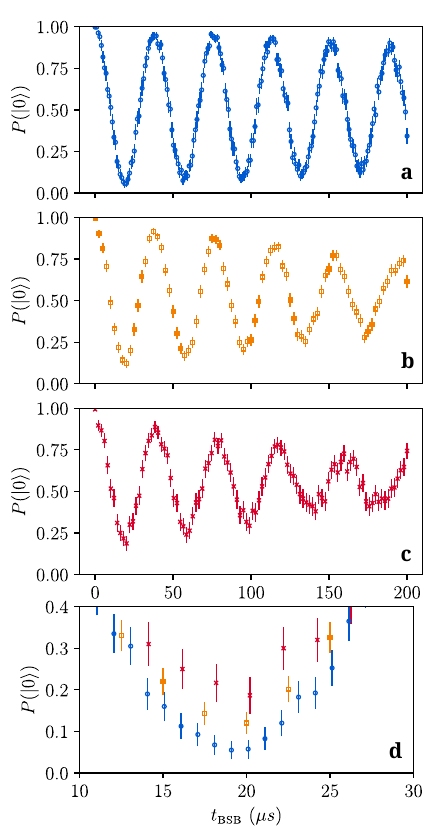}
  \caption{Blue-sideband oscillations after cooling. (a) A blue-sideband pulse is applied
    after 16 rounds of cooling have been performed on an initial thermal state with
    $\bar{n} \approx 15$. The blue-sideband pulse is followed by measuring the internal
    state of the ion by fluorescence detection. (b) Same as in (a) but after 17 rounds of
    cooling on an initial thermal state with $\bar{n} \approx 34$. (c) Same as in (a) but
    after 16 rounds of cooling on an initial thermal state with $\bar{n} \approx 51$. The
    plotted values are averages over 400 (a), 600 (b), 300 (c) experimental shots. Each
    plotted oscillation (a), (b), (c) corresponds to the cooling datasets plotted in
    Fig.~\ref{fig:ExpResults} of the main text. (d) Zoomed-in view with combined data from
    (a), (b), and (c) showing the first minima in the blue-sideband oscillations.}
  \label{fig:bsb-after-cool}
\end{figure}
To measure this effect we apply a blue-sideband pulse on the final cooled state after many
rounds of cooling and measure the internal ion qubit state. For a general oscillator state
$\rho = \sum_{j,k} \rho_{j k} \ket{j}_m\bra{k}_m$, where $\ket{n}_m$ are the oscillator
energy eigenstates, the probability to measure the internal qubit state $\ket{+Z}_s$ after
a blue-sideband pulse of duration $t$ on an initial state
$\ket{+Z}_s\bra{+Z}_s\otimes \rho$ is \cite{2003leibfried}
\begin{equation}
  \label{eq:bsb-fock}
  P_0(t) = \frac{1}{2}
  \left( 1 + \sum_{n=0}^\infty \rho_{nn} \cos\left(\Omega_{n,n+1} t\right) \right) \ ,
\end{equation}
where $\Omega_{n,n+1}$ are the Rabi frequencies of the blue-sideband interaction for the
transition between states $\ket{+Z}_s\ket{n}_m \leftrightarrow \ket{-Z}_s\ket{n+1}_m$. In
the Lamb-Dicke regime (for our experiments this requires $n < \sim 15$), $\Omega_{n,n+1}$
scales as $\sqrt{n+1}$. Thus we expect that the fraction of population in the ground state
leads to a sinusoidal oscillation of $P_0(t)$ while the tails at much higher energies lead
to a collection of higher frequency oscillations that quickly average out to $1/2$,
thereby reducing the contrast of the oscillation due to the ground state population. The
larger the fraction of population in the high energy tails, the lower the contrast of the
oscillation. In Fig.~\ref{fig:bsb-after-cool} we plot data measured after performing
blue-sideband pulses of varying duration, $t_{\mathrm{BSB}}$, after each of the cooling
sequences plotted in Fig.~\ref{fig:ExpResults}. We observe a sinusoidal oscillation due to
ground state population and a reduction of contrast as the temperature of the initial
thermal state before cooling increases. The rate of decay of the oscillations also
increases as the initial state temperature increases. This is expected due to the
increasing temperature of the radial modes, which have not been cooled, but whose
temperature increases proportionally to that of the initial state in the axial mode (which
we have cooled). As the temperature of the radial modes increases there is a broader
distribution of energies leading to a broader range of effective Rabi frequencies as given
by Eq.~\ref{eq:radial-Rabi-Fock}, leading to a faster decay of the oscillations plotted in
Fig.~\ref{fig:bsb-after-cool} as the temperature of the initial state is increased.

To estimate the fraction of population remaining in the tails of the Fock-state
distribution we fit a model that allows for exponential and Gaussian decay of the
oscillations given by Eq.~\ref{eq:bsb-fock}. When starting the cooling from thermal states
with $\bar{n} \approx$ 15 and 34, we extract a total population of $0.974 \pm 0.026$ and
$0.929 \pm 0.047$ respectively in the subspace spanned by $\ket{0}_m$ and $\ket{1}_m$ and
find negligible population estimates for Fock states $\ket{n}_m$, $n > 1$. After cooling
from an initial thermal state with $\bar{n} \approx 51$ we find a total population of
$0.873 \pm 0.177$ in Fock states with $n < 4$ and negligible populations in higher
levels. Thus, for the three datasets starting from thermal states with $\bar{n} \approx$
15, 34, 51 we estimate high energy tails with populations of $\sim 0.026$, $0.07$, and
$0.13$, respectively.

Although we observe high energy tails in the final state density matrices of the
simulations described above, we were unable to precisely predict the populations estimated
for the experiments using these simulations, which predict lower tail
populations. Furthermore, our simulations show that the tail populations can be reduced
even in the presence of all noise and imperfections simulated by performing slower cooling
using larger values of $\epsilon$ than optimal (see main text Fig.~\ref{fig:SemiClassical}
b)) and performing more cooling rounds. We have tried this in the experiment, but have not
observed a significant change in tail populations. Therefore, we conclude that the noise
model of the simulation is incomplete with respect to the experiments we have performed,
and further study would be required to fully understand the causes affecting tail
populations in these particular experiments. However, our simulations also demonstrate
that for any experiments well-described by the noise models we \emph{have} simulated, it
is possible to achieve much lower tail populations by performing somewhat slower cooling
with more rounds.

\section{Doppler cooling theory}

During Doppler cooling a drive laser excites a dipole allowed transition that decays in a
timescale $1/\Gamma$ due to spontaneous emission. For a drive with well-defined frequency
$\nu$ and atomic transition with frequency $\nu_0$, the probability of exciting the ion
with (dimensionless) momentum $p$, initially in the ground state, after some time $t$, is
given by \cite{2003leibfried, 2000loudon}
\begin{equation}
  P(-Z | p) = \frac{1}{\mathcal{N}} \frac{1}{\Gamma^2 + 4(\Delta - 2\eta\omega p)^2} \, ,
\end{equation}
where ${\Delta = \nu - \nu_0}$, ${\eta = \sqrt{\frac{\hbar k^2}{2m\omega}}}$ is the
Lamb-Dicke parameter, $k$ is the laser wavevector $2\pi/\lambda$ in the direction of the
ion motion, $m$ the ion mass, $\omega$ is the angular frequency of the ion motional
harmonic oscillator mode being cooled, and $\mathcal{N}$ is a normalisation constant to be
determined.

Close to the Doppler cooling limit we assume the amplitude of ion motion is small so that
the range of momentum $p$ over which the initial state probability is non-negligible is
such that the range of $4\eta\omega p$ is narrow compared to the natural linewidth of the
transition, $\Gamma$. In this case we can approximate $P(-Z|p)$ as a linear function in
$p$ (i.e. the Taylor expansion around $p=0$):
\begin{equation}
  \label{eq:lin-lorentz}
  P(-Z|p) \approx \frac{1}{\mathcal{N}}\frac{1}{\Gamma^2 + 4\Delta^2}
  \left( 1 + \frac{16\Delta}{\Gamma^2 + 4\Delta^2} \eta\omega p \right) \, .
\end{equation}
Given an ion with an initial thermal (i.e. Gaussian) distribution of momentum
\begin{align}
  f(p) = \frac{1}{\sqrt{2\pi}s} \mathrm{e}^{-p^2/ 2s^2} \, ,
\end{align}
where ${s = \sqrt{(\bar{n} + 1/2)/2}}$, we expect the constant $\mathcal{N}$ to be
time-dependent such that the probability $P(-Z|p)$ increases over time, since the last
emission. On average, a photon will be absorbed after a time $\tau$ such that the
probability of detecting the ion in the excited state, ${P(-Z) \equiv P_e}$. We can solve
for the value of $\mathcal{N}$ that satisfies this assumption, i.e.
\begin{equation}
  \int_{-\infty}^{\infty} P(-Z|p) f(p) dp = P(-Z) = P_e \, ,
\end{equation}
and using the linear approximation, Eq.~\ref{eq:lin-lorentz}, implies
\begin{equation}
  \label{eq:PmZ-N}
  \frac{1}{\mathcal{N}} = P_e \left( \Gamma^2 + 4\Delta^2 \right) \, .
\end{equation}
Then we have
\begin{equation}
  \label{eq:PmZ-normalised}
  P(-Z|p) \approx P_e \left( 1 + \frac{16\Delta}{\Gamma^2 + 4\Delta^2} \eta\omega p \right) \, .
\end{equation}

We can obtain an approximation of the energy after absorption and emission of a photon
using the semi-classical theory we have used to obtain Eq.~\ref{eq:meaneclass} in the main
text, but using Eq.~\ref{eq:PmZ-normalised}. That is, we assume an initially thermal state
with Gaussian probability density ${w_0(q,p) = f(q)f(p)}$ where we denote
${f(x) = \mathrm{e}^{-x^2 /(2s^2)}/(\sqrt{2\pi} s)}$, and ${s = \sqrt{(\bar{n}+1/2)/2}}$
as above. Neglecting the effect of spontaneous emission on the motion, the semi-classical
theory predicts a final probability density given by
\begin{equation}
  w_1(q,p) = f(q) \Big( P(+Z|p)f(p) + P(-Z|p - \eta) f(p - \eta) \Big) \, ,
\end{equation}
where ${P(+Z|p) = 1 - P(-Z|p)}$. The shift by $\eta$ for outcome $-Z$ is due to the
momentum transfer to the ion in photon absorption. No shift occurs in the absence of
photon absorption ($+Z$ outcome). The expected energy after a chosen time interval (which
determines the value of $P_e$) is
\begin{align}
  \langle E \rangle
  &= \hbar\omega \int_{-\infty}^{\infty} \int_{-\infty}^{\infty}
    w_1(q,p) \left( q^2 + p^2 \right) dq dp \nonumber \\
  &= 2\hbar\omega \left( \frac{1}{2} P_e \eta^2 + s^2
    \left(1 + P_e \eta^2 \frac{16 \omega \Delta}{\Gamma^2 + 4\Delta^2} \right) \right) \, .
    \label{eq:lin-lorentz-energy}
\end{align}

In the theory above we have not specified the probability of excitation $P_e$. If we
assume a constant drive field amplitude, we might expect a constant probability $\delta P$
per time step $\delta t$ since the last absorption-emission (at which point we know the
ion is in the ground state). i.e. the probability of excitation in a time $\delta t$ since
the last emission is $\delta P$, and the probability of excitation in a time window from
$(N-1)\delta t$ to $N\delta t$ is $(1 - \delta P)^{N-1} \delta P$, where $N$ is a positive
integer. In this case the probability density of excitation per unit time since the last
emission is of the form $f_e(t) = \frac{1}{\tau} \mathrm{e}^{-t/\tau}$. The average time
between absorption events is $\tau = \int_{0}^{\infty} f_e(t) t dt$ and the average
probability of excitation is
\begin{equation}
  \label{eq:ave-excitation-prob}
  P_e = \int_{0}^{\tau} f_e(t) dt = \frac{\mathrm{e} - 1}{\mathrm{e}} \, .
\end{equation}
Knowing the average time $\tau$ between absorption events, we could use
Eq.~\ref{eq:lin-lorentz-energy} with Eq.~\ref{eq:ave-excitation-prob} to estimate the
average cooling rate. Here we are mainly interested in calculating the minimum attainable
energy for which it happens that the value of $P_e$ chosen for the calculation is
irrelevant, as we now show.

Setting the final energy equal to the initial energy to find the steady-state energy,
i.e. $\langle E\rangle = 2\hbar\omega s^2$, we obtain (the factors of $P_e$ drop out)
\begin{equation}
  \label{eq:ss-nbar-linear-doppler}
  \bar{n} + \frac{1}{2} = 2s^2 = -\left( \frac{\Gamma^2 + 4\Delta^2}{16\omega\Delta} \right) \, .
\end{equation}
As in the standard Doppler cooling theory \cite{2003leibfried}, the energy is minimised by
a detuning $\Delta = -\Gamma/2$. This predicts a minimum energy corresponding to
\begin{equation}
  \label{eq:doppler-lim-linear-thy}
  \bar{n} + \frac{1}{2} = \frac{\Gamma}{4\omega} \, .
\end{equation}

We can relate the linear approximation of $P(\pm Z|p)$ to a linear approximation of
$P(\pm X|p)$ from the modular variable cooling theory, i.e.
\begin{equation}
  \label{eq:condprob-mom-Pe}
  P(\pm X | p) = P_e \left(1 \pm  \sin(4 \epsilon p) \right)
  \approx P_e \left( 1 \pm 4 \epsilon p \right) \, ,
\end{equation}
where ${P_e = 1/2}$ (see also Eq.~\ref{eq:condprob} in the main text). Using ${P_e = 1/2}$
in Eq.~\ref{eq:PmZ-normalised} and Eq.~\ref{eq:condprob-mom-Pe} matches the offsets. We
can then match the slope in Eq.~\ref{eq:condprob-mom-Pe} with that in
Eq.~\ref{eq:PmZ-normalised} by choosing
\begin{equation}
  \label{eq:eps-doppler}
  \epsilon = - \frac{4\eta\omega\Delta}{\Gamma^2 + 4\Delta^2} \, .
\end{equation}
Assuming $\epsilon^2s^2 \ll 1$, the energy predicted by Eq.~\ref{eq:meaneclass} with
$\epsilon$ given by Eq.~\ref{eq:eps-doppler} and $\alpha = \eta$ is approximately
\begin{equation}
  \label{eq:MV-cool-energy-doppler}
  \langle E\rangle_{\mathrm{classical}}
  \approx 2\hbar\omega
  \left( \eta^2 + s^2 \left( 1 + \eta^2
      \frac{32 \omega\Delta}{\Gamma^2 + 4\Delta^2} \right) \right)
  \, .
\end{equation}
As before, the cooling reaches a steady-state when the change in energy is zero, i.e.
$\langle E\rangle_{\mathrm{classical}} = 2\hbar\omega s^2$, which leads to
Eq.~\ref{eq:ss-nbar-linear-doppler}. Thus Eq.~\ref{eq:MV-cool-energy-doppler} predicts the
same minimum steady-state energy as Eq.~\ref{eq:doppler-lim-linear-thy}.

The Doppler limit expected from the standard theory for Doppler cooling, including
fluctuations due to photon absorption at different phases of the ion motion, but
neglecting the fluctuations due to spontaneous emission, is given by a temperature
\cite{2003leibfried}
\begin{equation}
  T_{\mathrm{min}} = \frac{\hbar \Gamma}{4 k_B} \, ,
\end{equation}
where $k_B$ is the Boltzmann constant. For a harmonic oscillator thermal state
\cite{2002barnett}
\begin{equation}
  \bar{n} = \frac{1}{\mathrm{e}^{\hbar\omega / k_B T} - 1} \, .
\end{equation}
When $\hbar\omega \ll k_B T$, we have
\begin{equation}
  \bar{n}_{\mathrm{min}} \approx \frac{k_B T_{\mathrm{min}}}{\hbar\omega} = \frac{\Gamma}{4 \omega} \, ,
\end{equation}
which matches Eq.~\ref{eq:doppler-lim-linear-thy} in the classical limit of the oscillator
state (i.e. $\bar{n} \approx \bar{n} + 1/2$; in any case the theory to reach
Eq.~\ref{eq:doppler-lim-linear-thy} assumes classical states).

\end{document}